\documentstyle [12pt,eqsecnum,aps] {revtex}
\input epsf
\newcommand{\beq}{\begin{equation}}
\newcommand{\eeq}{\end{equation}}
\newcommand{\beqs}{\begin{eqnarray}}
\newcommand{\eeqs}{\end{eqnarray}}

\begin{document}
\draft

\baselineskip 6.0mm

\title{Families of Graphs with $W_r(\{G\},q)$ Functions That Are Nonanalytic
at $1/q=0$} 

\author{Robert Shrock\thanks{email: shrock@insti.physics.sunysb.edu}
\and Shan-Ho Tsai\thanks{email: tsai@insti.physics.sunysb.edu}}

\address{
Institute for Theoretical Physics  \\
State University of New York       \\
Stony Brook, N. Y. 11794-3840}

\maketitle

\vspace{10mm}

\begin{abstract}

   Denoting $P(G,q)$ as the chromatic polynomial for coloring an $n$-vertex 
graph $G$ with $q$ colors, and considering the limiting function 
$W(\{G\},q) = \lim_{n \to \infty}P(G,q)^{1/n}$, a fundamental question in 
graph theory is the following: is $W_r(\{G\},q) = q^{-1}W(\{G\},q)$ analytic 
or not at the origin of the $1/q$ plane? (where the complex generalization of
$q$ is assumed).  This question is also relevant in statistical mechanics
because $W(\{G\},q)=\exp(S_0/k_B)$, where $S_0$ is the ground state entropy 
of the $q$-state Potts antiferromagnet on the lattice graph $\{G\}$, and the 
analyticity of $W_r(\{G\},q)$ at $1/q=0$ is necessary for the large-$q$ 
series expansions of $W_r(\{G\},q)$.  Although $W_r$ is analytic at
$1/q=0$ for many $\{G\}$, there are some $\{G\}$ for which it is not; for
these, $W_r$ has no large-$q$ series expansion.  It is important to 
understand the reason for this nonanalyticity.  Here we give a general 
condition that determines whether or not a particular $W_r(\{G\},q)$ 
is analytic at $1/q=0$ and explains the nonanalyticity where it occurs.  We 
also construct infinite families of graphs with $W_r$ functions that are 
non-analytic at $1/q=0$ and investigate the properties of these
functions. Our results are consistent with the conjecture that a sufficient
condition for $W_r(\{G\},q)$ to be analytic at $1/q=0$ is that $\{G\}$ is a 
regular lattice graph $\Lambda$.  (This is known not to be a necessary 
condition). 

\end{abstract}


\vspace{16mm}

\pagestyle{empty}
\newpage

\pagestyle{plain}
\pagenumbering{arabic}
\renewcommand{\thefootnote}{\arabic{footnote}}
\setcounter{footnote}{0}

\section{Introduction}

   The chromatic polynomial $P(G,q)$ of an $n$-vertex graph $G$ and the
asymptotic limiting function 
\beq
W(\{G\},q) = \lim_{n \to \infty} P(G,q)^{1/n}
\label{w}
\eeq
play important roles in both graph theory \cite{birk}-\cite{biggsbook} and 
statistical mechanics \cite{fk}-\cite{wurev}.  Here $P(G,q)$ is
defined as the number of ways of coloring the graph $G$ with $q$ colors such
that no two adjacent vertices have the same color, and $\{G\}$ denotes the
limit as $n \to \infty$ of the familiy of $n$-vertex graphs of type $G$. The 
connection with statistical mechanics is via the elementary equality 
$P(G,q) = Z(G,q,T=0)_{PAF}$, where $Z(G,q,T=0)_{PAF}$ is the 
partition function of the zero-temperature $q$-state Potts antiferromagnet (AF)
\cite{potts,wurev} on the graph $G$, and the consequent equality (in the $n \to
\infty$ limit) $W(\{G\},q)=\exp(S_0(\{G\},q)/k_B)$, where $S_0(\{G\},q)$ 
denotes the ground state entropy of the $q$-state Potts AF on $\{G\}$ 
(typically a regular lattice, $\{G\} = \Lambda$ with some specified boundary
conditions).  Given the fact that $P(G,q)$ is a polynomial,
there is a natural generalization, which we assume here, of the variable $q$ 
 from integer to complex values.  Since an obvious upper bound on $P(G,q)$
describing the coloring of an $n$-vertex graph with $q$ colors is 
$P(G,q) \le q^n$, and hence $W(\{G\},q) \le q$, it is natural to define a 
reduced function 
\beq
W_r(\{G\},q) = q^{-1}W(\{G\},q)
\label{wr}
\eeq
A fundamental question in graph theory concerns whether $W_r(\{G\},q)$ is
analytic or not at the origin, $1/q=0$, of the $z=1/q$ plane.  This question 
is important in both graph theory and statistical mechanics because a standard 
method for studying this function or equivalent reduced $W$ functions is to 
calculate a large-$q$ Taylor series expansion about the point $1/q=0$
\cite{nagle}-\cite{kewser}.  However, if $W_r(\{G\},q)$ is nonanalytic at 
$1/q=0$, then one cannot carry out such a Taylor series expansion in the 
usual manner.  Indeed, we recently discussed an example, namely, the 
bipyramid graph $B_n$, for which $W_r(\{B\},q)$ is not analytic at $1/q=0$ 
\cite{w} (see also \cite{read91}).  

   Clearly it is important to understand better the differences between 
the families of graphs that yield $W_r(\{G\},q)$ functions analytic at
$1/q=0$ and those which produce $W_r(\{G\},q)$ functions that are
nonanalytic at $1/q=0$.  In the present paper we shall address this problem.  
We shall give a general condition that determines whether or not a 
particular $W_r(\{G\},q)$ is analytic at $1/q=0$.  This explains the source of 
the nonanalyticity in the cases where it occurs.  We also construct 
infinite families of graphs with $W_r$ functions that are non-analytic at 
$1/q=0$.  These serve as a very useful theoretical laboratory, and we study
the properties of the resultant $W_r$ functions in some detail.  A salient 
point is that none of the $\{G\}$ that we construct with $W_r(\{G\},q)$ that 
are nonanalytic at $1/q=0$ is a regular lattice graph $\{G\}=\Lambda$.  Thus, 
anticipating the later discussion in this paper, our work is consistent with 
the conjecture that (in the $n \to \infty$ limit) a sufficient condition that 
$W_r(\{G\},q)$ be analytic at $1/q=0$ is that $\{G\}=\Lambda$ is a regular 
lattice graph.  (We know from our previous work \cite{w} that this is not a
necessary condition.)  We state it as a conjecture since we are not aware of 
any proof of the analyticity at $1/q=0$ of $W_r(\Lambda,q)$ for a regular 
lattice $\Lambda$ in the literature.  Indeed, in Ref. \cite{kewser}, it was 
acknowledged that there was no general theory for the existence of the limit 
(\ref{w}) and, hence also, in our notation, the reduced function $W_r$, even 
in the case of regular lattices.

   Before proceeding, it is 
necessary to clarify the definition of $W(\{G\},q)$ for values of
$q$ that are not positive integers.  As we discussed in Ref. \cite{w}, 
for certain ranges of real $q$, $P(G,q)$ can be negative, and, of course,
when $q$ is complex, so is $P(G,q)$ in general. In these cases it may not be
obvious, {\it a priori}, which of the $n$ roots
\beq
P(G,q)^{1/n} = \{ |P(G,q)|^{1/n}e^{2\pi i r/n} \} \ , \quad r=0,1,...,n-1
\label{pphase}
\eeq
to choose in eq. (\ref{w}).
Consider the function $W(\{G\},q)$ defined via eq. (\ref{w})
starting with $q$ on the positive real axis where $P(G,q) > 0$, and consider
the maximal region in the complex $q$ plane that can be reached by analytic
continuation of this function.  We denote this region as $R_1$.  Clearly, the
phase choice in (\ref{pphase}) for $q \in R_1$ is that given by $r=0$, namely
$P(G,q)^{1/n} = |P(G,q)|^{1/n}$. However, as we showed via exactly solved
cases in Ref. \cite{w}, there are many families of graphs $\{G\}$ for 
which the areas of analyticity of $W(\{G\},q)$ include other regions not
analytically connected to $R_1$, and in these regions, there is not, in
general, any canonical choice of phase in (\ref{pphase}).  

   A second subtlety in the definition of $W(\{G\},q)$ concerns the fact that
at certain special points $q_s$, the following limits do not commute \cite{w}
(for any choice of $r$ in eq. (\ref{pphase})):
\beq
\lim_{n \to \infty} \lim_{q \to q_s} P(G,q)^{1/n} \ne
\lim_{q \to q_s} \lim_{n \to \infty} P(G,q)^{1/n}
\label{wnoncomm}
\eeq
One can maintain the analyticity of $W(\{G\},q)$ at these
special points $q_s$ of $P(G,q)$ by choosing the order of limits in the
right-hand side of eq. (\ref{wnoncomm}):
\beq
W(\{G\},q_s)_{D_{qn}} \equiv \lim_{q \to q_s} \lim_{n \to \infty} P(G,q)^{1/n}
\label{wdefqn}
\eeq
As indicated, we shall denote this definition as $D_{qn}$, where the
subscript indicates the order of the limits.  Although this definition
maintains the analyticity of $W(\{G\},q)$ at the special points $q_s$, it
produces a function $W(\{G\},q)$ whose values at the points $q_s$ differ
significantly from the values which one would get for $P(G,q_s)^{1/n}$ with
finite-$n$ graphs $G$.  The definition based on the opposite order of limits,
\beq
W(\{G\},q_s)_{D_{nq}} \equiv \lim_{n \to \infty} \lim_{q \to q_s} P(G,q)^{1/n}
\label{wdefnq}
\eeq
gives the expected results like $W(G,q_s)=0$ for $q_s=0,1$, and, for 
$G \supseteq \triangle$, $q=2$, as well as $W((tri)_n,q=3)=1$ (where $(tri)_n$
denotes a triangular lattice with $n$ sites and boundary conditions that do not
introduce frustration for $q=3$), but yields a function $W(\{G\},q)$ with 
discontinuities at the set of points $\{q_s\}$.  
In our results below, in order to avoid having to write special
formulas for the points $q_s$, we shall adopt the definition $D_{qn}$ but at
appropriate places will take note of the noncommutativity of limits
(\ref{wnoncomm}).

\section{Construction of Families with 
$W_{\lowercase{r}}(\{G\},\lowercase{q})$ Nonanalytic at $1/\lowercase{q}=0$}
\label{construction}

\subsection{General Algorithm and Calculation of Chromatic Polynomial}

   In general, as discussed in Ref. \cite{w}, for a given family $\{G\}$, the 
corresponding $W(\{G\},q)$, at least as defined via the order of limits 
(\ref{wdefqn}), is an analytic function in certain regions of the complex $q$ 
plane.  These regions are separated from each other by curves (or lines)
comprising the union of boundaries ${\cal B}$. $W(\{G\},q)$ is nonanalytic
on these boundaries.  Clearly, ${\cal B}$ is the same for $W(\{G\},q)$ and 
$W_r(\{G\},q)$.  Because $P(G,q)$ is a polynomial with real
(actually integer) coefficients, it follows that  ${\cal B}$ is invariant under
complex conjugation, i.e., 
\beq
{\cal B}(q) = {\cal B}(q^*)
\label{binvariance}
\eeq
A basic question is whether, for a given family $\{G\}$,
some portion of the boundary ${\cal B}$ extends to complex infinity in the $q$
plane, so that $W_r(\{G\},q)$ is nonanalytic at $1/q=0$ in the $1/q$ plane. 
Related to this, an important question is whether there is a general 
algorithm for producing a family $\{G\}$ of graphs such that in the 
$n \to \infty$ limit, the boundary ${\cal B}$ extends to complex infinity in 
the $q$ plane.  We answer this question in the affirmative and present the 
following algorithm.  

   Consider a family of graphs $\{G\}$.  If this family already has the 
property that the limiting function $W(\{G\},q)$ has a region boundary 
${\cal B}$ that extends to complex infinity in the $q$ plane (i.e., to $1/q=0$
in the $1/q$ plane), then we have no 
work to do to get such a boundary. So assume that $\{G\}$ is such that 
$W(\{G\},q)$ has a region boundary ${\cal B}$ that does not extend to complex 
infinity in the $q$ plane.  As discussed in (section III and Theorem 1,
eq. (3.1) of) Ref. \cite{w}, a rather general form for the chromatic 
polynomial of a graph $G$ is 
\beq
P(G_n,q) = q(q-1)\Bigl \{  c_0(q) + \sum_{j=1}^{N_a} c_j(q)a_j(q)^n \Bigr \}
\label{pgsum}
\eeq
where $c_j(q)$ and $a_j(q)$ are polynomials in $q$. Here the $a_j(q)$ and
$c_{j \ne 0}(q)$ are independent of $n$, while $c_0(q)$ may contain 
$n$-dependent terms, such as $(-1)^n$, but does not grow with $n$ like $a^n$.
Obviously, the reality of $P(G,q)$ for real $q$ implies that $c_j(q)$ and 
$a_j(q)$ are 
real for real $q$.  The condition that ${\cal B}$ does not extend to infinite
distance from the origin in the $q$ plane is
equivalent to the condition that for sufficiently large $|q|$, there is one
leading term $a_j(q)$ in eq. (\ref{pgsum}).  Here we recall that ``leading
term $a_\ell(q)$ at a point $q$'' was defined in Ref. \cite{w} as a term 
satisfying $|a_\ell(q)| \ge 1$ and $|a_\ell(q)| > |a_j(q)|$ for $j \ne \ell$.
(If the
$c_0$ term is absent and $N_a=1$, then the sole $a_1(q)$ may be considered to
be leading even if $|a_j(q)| < 1$.)   We require sufficiently large $|q|$ so
that, for our analysis, there is a switching between only two leading terms
$a_\ell$.  In principle there might be such switching between more than two, so
that ${\cal B}$ would include more than two components running to complex 
infinity in the $q$ plane. However, for the families that we have constructed
via our algorithm and studied, we find, for sufficiently large $|Im(q)|$, 
only two such components.  As required by the symmetry (\ref{binvariance}), 
these components are mapped to each other under complex conjugation.  
Now adjoin a complete graph $K_p$ to $G_n$ in such a way that each vertex in
$K_p$ is adjacent, i.e., connected by bonds (=edges), to each of the
vertices of $G_n$.  Here, recall that a $p$-vertex graph is defined as 
``complete'' and labelled $K_p$ if each vertex is completely connected by 
bonds with all the other vertices of the $K_p$ graph. 
Denote the resultant graph as $(K_p \times G_n)$. 
A basic theorem of graph theory states that if a graph $H$ is 
obtained by adjoining a vertex to a graph $G$ such that this point is 
adjacent to all of the vertices of $G$, then the chromatic polynomials are
related according to
\beq
P(H,q) = qP(G,q-1)
\label{pgh}
\eeq
Applying this iteratively $p$ times, we obtain the result that
\beq
P(K_p \times G_n,q) = \Bigl [ \ \prod_{s=0}^{p-1}(q-s) \ \Bigr ]
P(G_n,q-p)
\label{pkg}
\eeq

   Next, we select one vertex in $K_p$ and remove $b$ bonds connecting it to
other vertices of $K_p$.  Since each vertex of $K_p$ has $p-1$ bonds connecting
it to other vertices of $K_p$, this implies that we can only remove this many
such bonds, i.e., 
\beq
1 \le b \le p-1
\label{bcondition}
\eeq
We denote the resultant graph as $(K_p \times G_n)_{rb}$, where the subscript 
signifies the above removal (r) of $b$ bonds.  In order for this to be
nontrivial, i.e., for $b \ge 1$, we thus require that 
\beq
p \ge 2
\label{pcondition}
\eeq
The conditions (\ref{pcondition}) and (\ref{bcondition}) will be 
assumed henceforth. Using eq. (\ref{pkg}) and $r$ applications of the 
addition-contraction theorem \cite{thm}, we obtain the important result 
\beqs
P((K_p \times G_n)_{rb},q) =  
P(K_p \times G_n,q) + b P(K_{p-1} \times G_n,q) \qquad \qquad \qquad \qquad 
\qquad \qquad \cr\cr
= \Bigl [ \ \prod_{s=0}^{p-2}(q-s) \ \Bigr ] 
\Bigl [ (q-(p-1))P(G_n,q-p) +b P(G_n,q-(p-1)) \Bigr ]
\label{pkpcutb}
\eeqs
This is our general formula for the chromatic polynomial of $(K_p \times
G_n)_{rb}$, for an arbitrary $n$-vertex graph $G_n$.  
Substituting the expression (\ref{pgsum}), we obtain 
\beqs
P((K_p \times G_n)_{rb},q) = \Bigl [ \ \prod_{s=0}^{p-2}(q-s) \ \Bigr ]
\times \qquad \qquad \qquad \qquad \qquad \qquad \qquad \qquad \qquad \cr 
\biggl [ (q-p+1)(q-p)(q-p-1) \Big \{ c_0(q-p) + 
\sum_{j=1}^{N_a} c_j(q-p)a_j(q-p)^n \Big \} \cr
+ b(q-p+1)(q-p) \Big \{ c_0(q-p+1) + 
\sum_{j=1}^{N_a} c_j(q-p+1)a_j(q-p+1)^n \Big \} \biggr ]
\label{pkpcutbfull}
\eeqs

\subsection{Boundary ${\cal B}$ for $\{(K_p \times G)_{rb}\}$}

   We denote the $n \to \infty$ limit of the families $K_p \times G_n$ and
$(K_p \times G_n)_{rb}$ as 
\beq
\lim_{n \to \infty} K_p \times G_n = \{K_p \times G\}
\label{kpginf}
\eeq
and
\beq
\lim_{n \to \infty}(K_p \times G_n)_{rb} = \{(K_p \times G)_{rb}\}
\label{kpgr1inf}
\eeq
respectively.  As discussed in Ref. \cite{w}, the boundary ${\cal B}$ for 
$W(\{G\},q)$ is the locus of points
in the $q$ plane where there is a switching between different leading terms
$a_\ell$ in eq. (\ref{pgsum}).  Since $\{G\}$ was assumed not to have 
${\cal B}$ extending to infinity in the $q$ plane, it follows that for large 
enough $|q|$, there is a single leading term $|a_\ell(q)|$ in
(\ref{pgsum}). Hence from eq. (\ref{pkpcutbfull}), we see that at 
sufficiently large $|q|$, the boundary ${\cal B}$ for the limiting function 
$W(\{K_p \times G)_{rb} \},q)$ is determined by the equality
\beq
|a_\ell(q-p)| = |a_\ell(q-p+1)|
\label{mageq}
\eeq
Note that this is independent of $b$, so that
\beq
{\cal B} \quad {\rm is \ \ independent \ \ of} \quad b \quad {\rm for} \quad
\{(K_p \times G)_{rb}\}
\label{bindb}
\eeq
(given that the basic condition (\ref{bcondition}) is satisfied). 
Let $q = q_{_R} + iq_{_I}$.  One can next enumerate the various cases 
possible for $a_\ell(q)$.  The basic theorem that the coefficient of the
highest-order term, $q^n$, in the chromatic polynomial $P(G_n,q)$ of any
$n$-vertex graph $G_n$ is unity implies that if a dominant term $a_\ell(q)$ 
is a polynomial of degree $s_{max}$,
\beq
a_\ell(q) = \sum_{s=0}^{s_{max}} \alpha_s q^s
\label{apolynomial}
\eeq
then 
\beq
\alpha_{s_{max}}=1
\label{alphasmax}
\eeq
Consider, for example, the case where $a_\ell(q)$ is a linear function of
$q$: $a_\ell(q) = \alpha_1 q + \alpha_0$, which reduces to 
$a_\ell(q) = q + \alpha_0$ by (\ref{alphasmax}).  Then (\ref{mageq}) yields 
\beq
q_{_R} = p - \Big ( \frac{1}{2} + \alpha_0 \Big ) \quad {\rm for} \quad 
s_{max}=1
\label{qrsol}
\eeq
with $q_{_I}$ undetermined, i.e., a vertical line segment extending to 
$\pm i \infty + p -(\frac{1}{2}+\alpha_0)$ in the complex $q$ plane.  
This type of behavior is exemplified by graphs involving $K_p$ adjoined
to trees or chains of triangles, in which one bond in the $K_p$ subgraph is
removed. We shall discuss these below. 

If $a_\ell(q)$ is a quadratic function of $q$, 
$a_\ell(q) = q^2 + \alpha_1 q + \alpha_0$, then (\ref{mageq}) yields an 
equation which has, as its only acceptable solution,
\beq
q_{_R} = p -\frac{1}{2}(1+\alpha_1) \quad
{\rm as} \quad |q_I| \to \infty \quad {\rm for} \quad s_{max}=2
\label{qrsolquad}
\eeq
Hence, the boundary ${\cal B}$ in this case is, for sufficiently large $|q_I|$,
again a vertical line in the complex $q$ plane located at the value of $q_{_R}$
given by (\ref{qrsolquad}) and extending to $\pm i \infty$. 
This type of behavior is exemplified in our discussion below of graphs
involving $K_p$ adjoined to chains of squares (i.e., ladder graphs) with 
various boundary conditions, in which $r$ bonds are removed from the $K_p$
subgraph in the manner discussed above. 
In general, as we shall show, if one adjoins $K_p$ to an open chain of $k$-gons
arranged such that two adjacent $k$-gons intersect along one of their mutual
edges, then the resultant chromatic polynomial has the form (\ref{pkpcutbfull})
with $s_{max}=k-2$ and $c_0(q)=0$ (there should be no confusion in the notation
of $k$ for the $k$-gons and $K$ for $K_p$). 

    This, then, is the algorithm for producing families of graphs depending on
three parameters, $p$, $b$, and $n$, with the property that the limiting 
function $W_r$ is nonanalytic at $1/q=0$.  We have proved this by calculating 
first the chromatic polynomials for finite graphs and then their respective 
limiting functions $W$.  The key ingredients in the construction are, first, 
the adjoining of the complete graph $K_p$ to $G_n$, and second, the removal 
of $b$ of the bonds connecting one vertex in $K_p$ to other vertices of $K_p$.
Together, these guarantee, via eq. (\ref{pkpcutb}), that the equation for the 
degeneracy of the leading term $a_\ell$ is of the form (\ref{mageq}), and the 
locus of points that solve this degeneracy equation extends to complex 
infinity in the $q$ plane, as we have shown.  

   We next give some illustrations of the application of this algorithm. 

\section{Families of Graphs with $W_{\lowercase{r}}$ Nonanalytic at 
$1/\lowercase{q}=0$}

\subsection{ $(K_p \times T_n)_{rb}$ }

   Perhaps the simplest illustration is for the family $K_p \times T_n$
formed by adjoining a succession of $p$ vertices to each of the $n$ vertices
of a tree graph $T_n$ and to each other.  Removing $b$ bonds connecting a
vertex of $K_p$ to other vertices of $K_p$ in the manner described above yields
the graph $(K_p \times T_n)_{rb}$. From eq. (\ref{pkpcutb}), we calculate 
\beq
P((K_p \times T_n)_{rb},q) = \Bigl [ \prod_{s=0}^{p}(q-s) \Bigr ]
     \Bigl [ (q-p-1)^{n-1} + b(q-p)^{n-2} \Bigr ]
\label{pkptrb}
\eeq
Our general analysis in eqs. (\ref{mageq})-(\ref{qrsol}) applies with
$s_{max}=1$, $\alpha_0=-1$, and $c_0(q)=0$ so that, from the general 
formula (\ref{qrsol}) we have 
\beq
q_{_R} = p+ \frac{1}{2} \quad {\rm for} \quad G = (K_p \times T)_{rb}
\label{qrkpt}
\eeq
Hence, the boundary ${\cal B}$ consists of the vertical line (\ref{qrkpt}) with
$-i \infty \le q_{_I} \le i\infty$.  The diagram 
describing the regions of analyticity of the limiting function 
$W(\{(K_p \times T)_{rb}\},q)$ consists of two regions, 
\beq
R_1 : Re(q) > p + \frac{1}{2} 
\label{bkptr1}
\eeq
and 
\beq
R_2 : Re(q) < p + \frac{1}{2}
\label{bkptr2}
\eeq
Mapped to the $1/q$ plane (a conformal transformation), the image of the
vertical line is a closed curve which crosses the real axis at the inverse of
$q_{_R}$ in (\ref{qrkpt}) and at the origin.  In the $1/q$ plane, the image 
of region $R_1$ is a compact region enclosed by this closed curve, while its 
complement is the image of the region $R_2$.  We find that 
\beq
W(\{(K_p \times T)_{rb}\},q) = q-p \quad {\rm for} \quad q \in R_1
\label{wbkptr1}
\eeq
i.e., $W_r(\{(K_p \times T)_{rb}\},q)=1-p/q$. 
For $q \in R_2$, if $q$ is real, $P((K_p \times T_n)_{rb},q)$ alternates in 
sign as $n$ increases through even and odd integers, so, strictly speaking, 
the limit as $n \to \infty$ does not exist.  Of course, there is also a
corresponding variation in phases in this limit for the case of complex 
$q \in R_2$.  As we have discussed before, in such a situation, at
least the magnitude does have a well-defined limit:
\beq
|W(\{(K_p \times T)_{rb}\},q)| = |q-p-1| \quad {\rm for} \quad q \in R_2
\label{wbkptr2}
\eeq
or equivalently $|W_r(\{(K_p \times T)_{rb}\},q)| = |1-\frac{p+1}{q}|$. 
This simple example thus explicitly illustrates the nonanalyticity at
$1/q=0$; even aside from the choice of the phase in region $R_2$, the two
expressions above for the magnitude of the reduced $W_r$ functions are 
different.  The nonanalyticity in the $W_r$ function thus involves (for any
choice of phases in (\ref{pphase}) in $R_2$) a discontinuity in the first
derivative $dW_r/dz$, where $z=1/q$ at $z=0$.  

   Although these are all exact results for the $n \to \infty$ limit, it is of
some interest to 
see how these boundaries develop by studying chromatic zeros for finite $n$.
(Here, we recall the definition that the ``chromatic zeros'' of a graph $G$ 
are the zeros of the chromatic polynomial $P(G,q)$ for this graph.)
We carried out these types of studies for a number of families of graphs in
Refs. \cite{w} and \cite{wc}.   A general question that we investigated was 
the following: excluding a well-understood subset of chromatic zeros at 
certain discrete real integer values, and considering the remainder, how close
are these remaining zeros, for finite $n$, to the boundary ${\cal B}$ that 
obtains in the $n \to \infty$ limit?  From our study of the bipyramid graph,
in Ref. \cite{w} (see also Ref. \cite{read91}), it was found that the 
chromatic zeros near to the real axis in the $q$ plane (aside from the 
discrete zeros at $q=0$ and 1 and a zero very near to $q=Be_5 = 2.618...$) 
lie on or near to the arcs forming the boundary curves ${\cal B}(R_1,R_3)$ 
and ${\cal B}(R_2,R_3)$ (defined in greater generality in eqs. (\ref{leftarc}),
(\ref{rightarc}) below), but the outer zeros do not lie very close to the 
line segments of ${\cal B}(R_1,R_2)$ (given by the $p=2$ special case of eq. 
(\ref{lines}) below) and only approach these line segments slowly as $n$ 
increases.  We inferred that this latter behavior was connected with the fact 
that this component of the boundary extends to complex infinity in the $q$
plane and observed that this type of deviation did not occur for families of
graphs whose boundaries ${\cal B}$ were compact and did not extend to complex
infinity. 

\begin{figure}
        \centering
        \leavevmode
\epsfxsize=3.5in
\epsffile{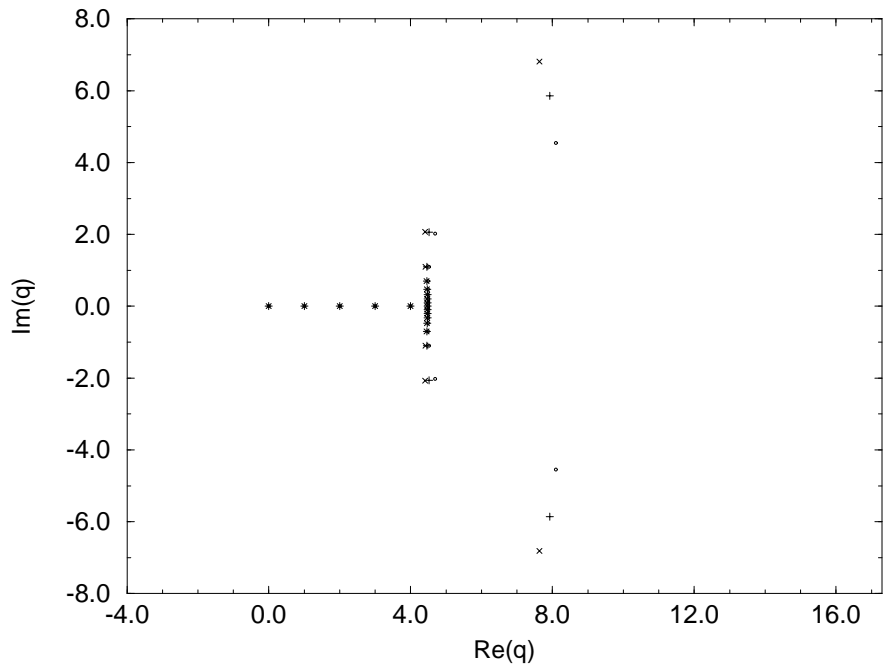}
\caption{Chromatic zeros of $(K_p \times T_n)_{rb}$ for $p=4$, $n=18$, 
and $b=1$ ($\cdot$), 2 ($+$), and 3 ($\times$), where the symbols for the
points are given in parentheses. In the $n \to \infty$ limit, our exact results
give the continuous region boundary ${\cal B}$ as the $p=4$ special case of 
eq. (\ref{qrkpt}), i.e., the vertical line with $Re(q)=9/2$ and $Im(q)$
arbitrary, independent of $b$ by eq. (\ref{bindb}).}
\label{k4tnrbfig}
\end{figure}

    Here we have carried out an analogous study of the chromatic zeros of 
$(K_p \times T_n)_{rb}$ and we find similar behavior as a function of $n$.  
Since an interesting feature here is the dependence of the locations of 
chromatic zeros on $b$, we focus on this. In Fig. \ref{k4tnrbfig} we show the 
results for $p=4$ and $n=18$ for $b=1,2$, and 3, the full range allowed by 
(\ref{bcondition}). The outermost zero and its complex conjugate do not lie 
very near to the vertical line with $q_{_R}=9/2$ given by the $p=4$ special 
case of eq. (\ref{qrkpt}).  As $b$ increases from 1 to 3, this 
outermost zero moves to larger $|q_I|$ and slightly smaller $q_{_R}$, and hence
slowly toward the above-mentioned vertical line.  From eq. (\ref{pkptrb}), it
follows that for $p=4$, there are also discrete chromatic zeros at $q=0,1,2,3$,
and 4, and these are evident in Fig. \ref{k4tnrbfig}.  

\subsection{ $(K_p \times C_n)_{rb}$ }

\begin{figure}
\epsfxsize=3.5in
\epsffile{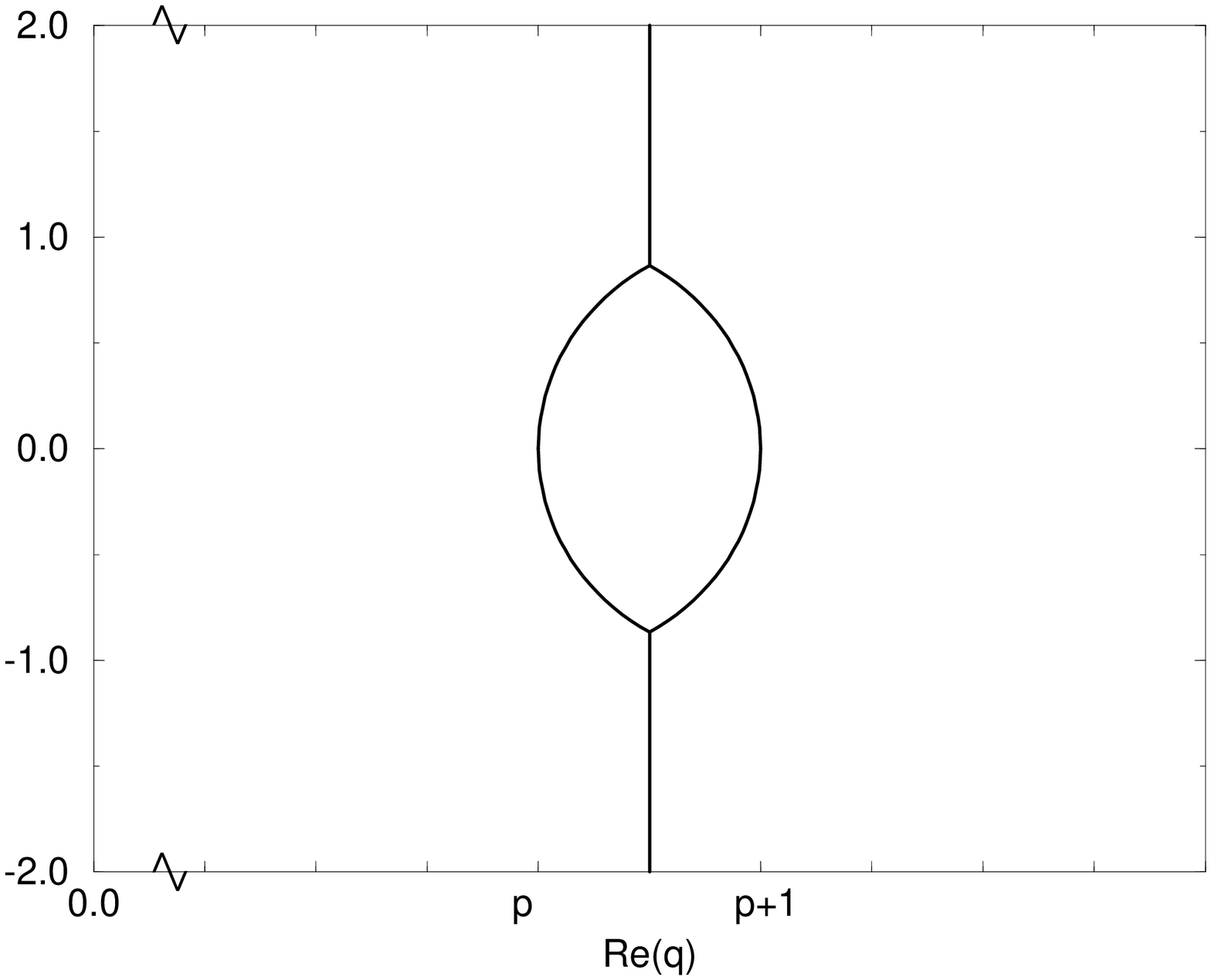}
\caption{Diagram showing regional boundaries comprising ${\cal B}$ for
$W(\{(K_p \times C)_{rb}\},q)$. Breaks in the horizontal axis indicate that
$p$ is an arbitrary integer $\ge 2$.}
\label{kptrbfig}
\end{figure}

    A second illustration is provided by the family of graphs 
$(K_p \times C_n)_{rb}$ obtained by starting with the $p$-wheel 
$K_p \times C_n$ with $p \ge 2$, where $C_n$ is the $n$-vertex circuit graph, 
and removing $b$ bonds connecting a vertex in $K_p$ to other vertices in 
$K_p$.  From the general formula eq. (\ref{pkpcutb}) we calculate
\begin{eqnarray}
\lefteqn{
P((K_p \times C_n)_{rb},q) = \Bigl [ \prod_{s=0}^{p-2}(q-s) \Bigr ] \times} \\ 
\nonumber 
& & \times 
\biggl [  (q-p+1)(q-p-1) \Bigl \{ (q-p-1)^{n-1} + (-1)^n \Bigr \} 
+ b(q-p) \Bigl \{ (q-p)^{n-1} + (-1)^n \Bigr \} \biggr ]
\label{pkpcrb}
\end{eqnarray}
Our general analysis in eqs. 
(\ref{mageq})-(\ref{qrsol}) applies with $s_{max}=1$ and $\alpha_0=-1$ so 
that again, for sufficiently large $|q|$, ${\cal B}$ contains the vertical
lines extending to $\pm i \infty$ with $q_{_R} = p+1/2$, as was the case with 
$W(\{(K_p \times T)_{rb}\},q)$.  
However, because the $c_0(q)$ terms are nonzero in
this case, the region boundaries in the vicinity of the real axis differ from
the simple vertical line found for $W(\{(K_p \times T)_{rb}\},q)$.  By methods
similar to those which we used in Ref. \cite{w} (i.e. working out the
conditions for the degeneracy of the leading terms in $P$), we find that the 
region diagram for $W(\{(K_p \times C)_{rb}\},q)$ consists of three regions: 
\beq
R_1: Re(q) > p + \frac{1}{2} \quad {\rm and} \quad |q-p| > 1
\label{br1}
\eeq
\beq
R_2: Re(q) < p + \frac{1}{2} \quad {\rm and} \quad |q-(p+1)| > 1
\label{br2}
\eeq
and
\beq
R_3: |q-p| < 1 \quad {\rm and} \quad |q-(p+1)| < 1
\label{br3}
\eeq
The boundaries between these regions are thus the two circular arcs
\beq
{\cal B}(R_1,R_3):
q = p+e^{i\theta} \ , -\frac{\pi}{3} < \theta < \frac{\pi}{3}
\label{leftarc}
\eeq
and
\beq
{\cal B}(R_2,R_3): q=p+1+e^{i\phi} \ , \frac{2\pi}{3} < \phi < \frac{4\pi}{3}
\label{rightarc}
\eeq
together with the semi-infinite vertical line segments
\beq
{\cal B}(R_1,R_2) = \{q\}: \quad
Re(q)=p + \frac{1}{2} \quad {\rm and} \quad |Im(q)| > \frac{\sqrt{3}}{2}
\label{lines}
\eeq
These meet at the intersection points
\beq
q_{int.} = p + \frac{1}{2} \pm i\frac{\sqrt{3}}{2}
\label{qint}
\eeq 
The arcs ${\cal B}(R_1,R_3)$ and ${\cal B}(R_2,R_3)$ cross the real axis at
$q=p+1$ and $q=p$, respectively.  The region diagram for 
$W(\{(K_p \times C)_{rb}\},q)$ is shown in Fig. \ref{kptrbfig}. 

Mapped to the $1/q$ plane, the image of the vertical line is a curve in the
right-hand half-plane which passes vertically through the origin, enclosing the
image of $R_1$.  The enclosed region $R_3$ remains enclosed, while the
left-hand region $R_2$ in the $q$ plane maps to the region exterior to the
images of $R_1$ and $R_3$ in the $1/q$ plane. 
The case $p=2$ is the bipyramid family of graphs \cite{w},\cite{read91}.
We find that 
\beq
W(\{ (K_p \times C)_{rb} \},q) = q-p \quad {\rm for} \ \ q \in R_1
\label{wbr1}
\eeq
For the other regions, we have, in general,
\beqs
|W(\{ (K_p \times C)_{rb} \},q)| & = & |q-(p+1)| \quad {\rm for} \ \ 
q \in R_2 \cr
               & = & 1     \quad {\rm for} \ \ q \in R_3
\label{wb}
\eeqs
The $W_r$ functions in regions $R_1$ and $R_2$ are the same as those in the
corresponding regions for the previous family, $\{(K_p \times T)_{rb}\}$, while
region $R_3$ has no analogue for that family. 
As before, the nonanalyticity in $W_r$ is most conveniently discussed in the
$1/q$ plane.  There are two regions contiguous across the image of ${\cal B}$
at the origin, $1/q=0$, namely, the images under inversion of regions $R_1$ and
$R_2$.  The nature of the nonanalyticity in $W_r$ at $1/q=0$ is the same as 
that in the $\{(K_p \times T)_{rb}\}$ family. 

\begin{figure}
\epsfxsize=3.5in
\epsffile{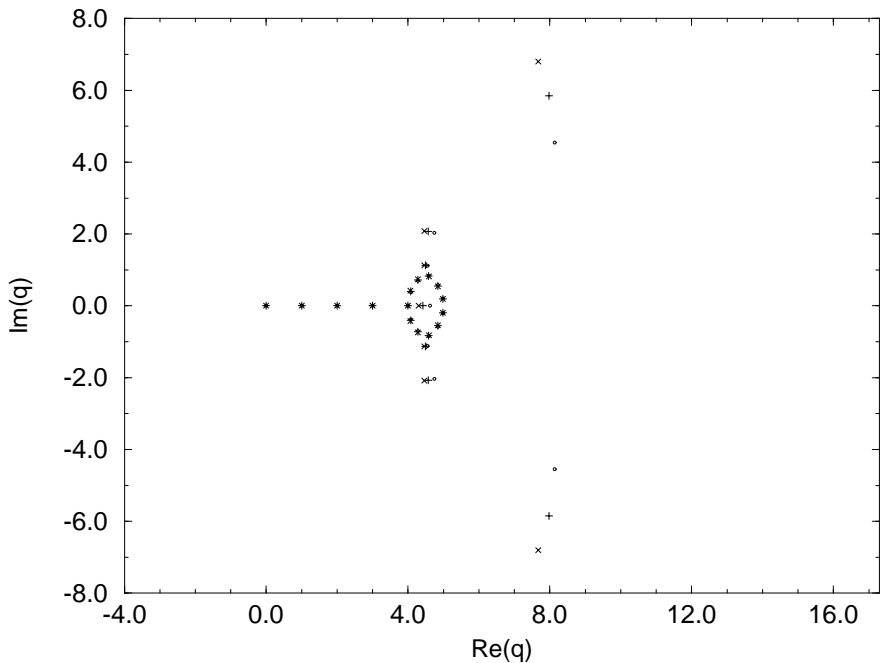}
\caption{Chromatic zeros of $(K_p \times C_n)_{rb}$ for $p=4$, $n=18$,
and $b=1$ ($\cdot$), 2 ($+$), and 3 ($\times$).  In the $n \to \infty$ limit, 
our exact results give the continuous region boundary ${\cal B}$ as the $p=4$ 
special case of Fig. \ref{kptrbfig}.}
\label{k4cnrbfig}
\end{figure}

   Again, it is of interest to calculate the chromatic zeros for some finite
graphs in these families and see how close they lie to the locus of zeros in
the $n \to \infty$ limit.  We have carried out such a study. As an
illustration, in Fig. \ref{k4cnrbfig}, we show the chromatic zeros for 
$(K_p \times C_n)_{rb}$ with $p=4$, and $n=18$ for $b=1,2$ and 3. 
This may be compared with the $p=4$
special case of the plot of the region diagram in Fig. \ref{kptrbfig} (which is
the same for all $b$, by the result (\ref{bindb})).  Note 
that the discrete chromatic zeros at $q=0,1,2$, and 3 are not part of the 
continuous locus of zeros forming ${\cal B}$ in the $n \to \infty$ limit. 
For comparison, the previously mentioned
bipyramid graph is the case $p=2$, for which only the single value $b=1$ is 
allowed.  We find that the complex chromatic 
zeros near the real axis lie on or close to the arcs 
${\cal B}(R_1,R_3)$ and ${\cal B}(R_2,R_3)$, but the outer zeros do not lie 
very close to the line segments of ${\cal B}(R_1,R_2)$ given by the 
$p=4$ special case of (\ref{lines}), namely, $q_{_R}=9/2$, 
$|q_I| \ge \sqrt{3}/2$ and only approach these line segments slowly as $n$ 
increases.  Evidently, the behavior of the chromatic zeros as
functions of $b$ is qualitatively similar to that which we observed in Fig. 
\ref{k4tnrbfig}.  We have carried out analogous studies of the chromatic zeros
for other families of graphs constructed via our algorithm to have region 
boundaries ${\cal B}$ extending to complex infinity and have found similar 
results. 

\subsection{$(K_p \times (Ch)_{k,n})_{rb}$}

   We define an open chain of $m$ $k$-gons constructed such that a given 
$k$-gon intersects with the next $k$-gon in the chain along one of their 
mutual edges as $(Ch)_{k,n}$ where the number of vertices is
\beq
n = (k-2)m+2
\label{chmn}
\eeq
The chromatic polynomial for the open chain of $m$ $k$-gons is 
\beq
P((Ch)_{k,n},q) = q(q-1)D_k(q)^m
\label{pchaingkgons}
\eeq
where
\beq
D_k(q) = \sum_{s=0}^{k-2}(-1)^s {{k-1}\choose {s}} q^{k-2-s} 
\label{dk}
\eeq
Next, consider the graph $K_p \times (Ch)_{k,n}$ obtained by adjoining $K_p$ 
to the chain of $m$ $k$-gons, $(Ch)_{k,n}$, where $n$ is again given by
eq. (\ref{chmn}).  From (\ref{pkg}), we have 
\beq
P(K_p \times (Ch)_{k,n},q) = \Bigl [\prod_{s=0}^{p+1}(q-s)\Bigr ] D_k(q-p)^m
\label{pkpch}
\eeq
The chromatic polynomial $P((K_p \times (Ch)_{k,n})_{rb},q)$ is then obtained 
by substitution of (\ref{pkpch}) into eq. (\ref{pkpcutb}). Note that 
\beq
D_k(q)=a_\ell(q)
\label{dka}
\eeq
in eqs. (\ref{pkpcutbfull}), (\ref{mageq}), and (\ref{apolynomial})
general class of graphs.  From the above substitution, we find that the 
resultant chromatic polynomial is of the form (\ref{pkpcutbfull}) with 
\beq
c_0(q)=0 \ , \quad s_{max}=k-2 \quad {\rm for} \quad 
(K_p \times (Ch)_{k,n})_{rb}
\label{csmax}
\eeq
For sufficiently large $|q_I|$, the region boundary ${\cal B}$ for 
$W(\{(K_p \times (Ch)_k)_{rb}\},q)$ consists of complex-conjugate vertical 
line segments extending to $\pm i \infty$ with $Re(q) = q_{_R}$ equal to 
\beq
q_{_R} = p+\frac{k}{2(k-2)} \quad {\rm for} \quad \{(K_p \times (Ch)_k)_{rb}\}
\quad {\rm with} \quad k \ge 3
\label{qrkgonchain}
\eeq
where the notation convention of (\ref{kpgr1inf}) is used.  Note that if one
considers chains of $k$-gons with $k$ progressively larger and larger, this
approaches the limit $\lim_{k \to \infty} q_{_R} = p + 1/2$.  As defined in the
Introduction, the region to the right of this boundary is $R_1$, and we denote
the region to the left as $R_2$.  In general (in particular, for $k \ge 5$)
there will be other regions (within which $W(\{(K_p \times (Ch)_k)_{rb}\},q)$
is analytic) but we only need $R_1$ and $R_2$ for our discussion of the
nonanalyticity at $1/q=0$ since it is only the images under inversion of 
these two regions that border the origin of the $1/q$ plane. 
For the resultant limiting functions $W$ we calculate
\beq
W(\{(K_p \times (Ch)_k)_{rb}\},q) = D_k(q-p+1)^{\frac{1}{k-2}} \quad 
{\rm for} \quad q \in R_1
\label{wkgonchainr1}
\eeq
\beq
|W(\{(K_p \times (Ch)_k)_{rb}\},q)| = |D_k(q-p)|^{\frac{1}{k-2}} \quad 
{\rm for} \quad q \in R_2
\label{wkgonchainr2}
\eeq
Hence 
\beq
W_r(\{(K_p \times (Ch)_k)_{rb}\},q) = \biggl (1 - \frac{p-1}{q} \biggr ) 
\Bigl [ \sum_{s=0}^{k-2}(-1)^s {{k-1} \choose {s}} (q-p+1)^{-s} 
\biggr ]^{\frac{1}{k-2}} \quad {\rm for} \quad q \in R_1
\label{wrkgonchainr1}
\eeq
\beq
|W_r(\{(K_p \times (Ch)_k)_{rb}\},q)| = \biggl | \biggl (1-\frac{p}{q}\biggr ) 
\Bigl [ \sum_{s=0}^{k-2}(-1)^s {{k-1} \choose {s}} (q-p)^{-s} \Bigr ] 
\biggr |^{\frac{1}{k-2}} \quad {\rm for} \quad q \in R_2
\label{wrkgonchainr2}
\eeq
For $k \ge 4$, these have the expansions near $1/q=0$ 
\beq
W_r(\{(K_p \times (Ch)_k)_{rb}\},q) = 1 - \biggl ( p-1 + \frac{k-1}{k-2} 
\biggr ) q^{-1} + O(q^{-2}) 
\quad {\rm as} \quad 1/q \to 0 \quad {\rm with} \quad q \in R_1
\label{wrkgonchainr1taylor}
\eeq
(e.g., $1/q \to 0^+$ through real values) and
\beq
|W_r(\{(K_p \times (Ch)_k)_{rb}\},q)| = \biggl | 1 - 
\biggl (p + \frac{k-1}{k-2} \biggr )q^{-1} + O(q^{-2})
\biggr | \quad {\rm as} \quad 1/q \to 0 \quad {\rm with} \quad q \in R_2
\label{wrkgonchainr2taylor}
\eeq
These expressions also apply for the case $k=3$ but terminate with the
$O(q^{-1})$ term, since $D_3(q)$ is linear and the exponent $1/(k-2)$ in 
eqs. (\ref{wkgonchainr1}) and (\ref{wkgonchainr2}) is just unity.  Thus, for
all $k \ge 3$, these results indicate explicitly the nature of the
nonanalyticity of the reduced function $W_r(\{(K_p \times (Ch)_k)_{rb}\},q)$ at
$1/q=0$.  As in our earlier examples, this nonanalyticity involves, in general,
the sudden onset of a complex phase and apart from this, even for the
magnitude, a discontinuity in the first derivative $dW_r/dz$ at $z=1/q=0$. 

   Since the lowest two cases, $k=3,4$ exhibit particularly simple boundaries
${\cal B}$, we give a few explicit results for them. For $k=3$ we find 
\beq
P((K_p \times (Ch)_{3,n})_{rb},q) = \Bigl [ \prod_{s=0}^p (q-s) \Bigr ]
\Bigl [ (q-p-1)(q-p-2)^m + b(q-p-1)^m \Bigr ]
\label{pkpcutbchain3gon}
\eeq
where from (\ref{chmn}), $m=n-2$.  This has the form of our
general eq. (\ref{pkpcutbfull}) with linear $a_\ell(q)$, $\alpha_0=-2$ and 
$c_0(q)=0$ so that in the $n \to \infty$ limit, by (\ref{qrsol}), it follows
that the boundary ${\cal B}$ consists of the vertical line 
\beq
q_{_R} = p + \frac{3}{2} \quad {\rm for} \quad \{(K_p \times (Ch)_3)_{rb}\}
\label{qrtrianglechain}
\eeq
The general formulas (\ref{wrkgonchainr1}) and (\ref{wrkgonchainr2}) reduce to
the simple expressions 
\beq
W_r(\{(K_p \times (Ch)_3)_{rb}\},q) = 1-\frac{p+1}{q} \quad {\rm for} \quad q 
\in R_1
\label{wr3gonchainr1}
\eeq
and 
\beq
|W_r(\{(K_p \times (Ch)_3)_{rb}\},q)| = \biggl | 1 - \frac{p+2}{q} \biggr | 
\quad {\rm for } \quad q \in R_2
\label{wr3gonchainr2}
\eeq

For the $k=4$ case, $D_4(q)=a_\ell(q)=q^2 -3q + 3$, 
so that (\ref{qrkgonchain}) or the quadratic special case eq. (\ref{qrsolquad})
with $\alpha_1=-3$ and $c_0(q)=0$ applies and yields 
\beq
q_{_R} = p + 1 \quad {\rm for} \quad W(\{(K_p \times (Ch)_4)_{rb}\},q)
\label{qr4gonchain}
\eeq
Using (\ref{pkpch}) and (\ref{pkpcutb}), we determine that the region boundary
of $W(\{(K_p \times (Ch)_4)_{rb}\},q)$ is again precisely the vertical line
with $q_{_R}$ given by (\ref{qr4gonchain}) and $|q_I|$ arbitrary.  Note that
this holds even though $D_4(q)=a_\ell(q)$ is quadratic. However, the generic
behavior for higher $k$ is that in the vicinity of the real axis, ${\cal B}$ is
more complicated, and the complex-conjugate vertical line segments with 
$q_{_R}$ given by (\ref{qrkgonchain}) apply for $|q_I| > \kappa_k$, where
$\kappa_k$ is a $k$-dependent constant.  

   From eqs. (\ref{wrkgonchainr1}) and (\ref{wrkgonchainr2}) we have
\beq
W_r(\{(K_p \times (Ch)_4)_{rb}\},q) = \Bigl [1 -(2p+1)q^{-1}+(p^2+p+1)q^{-2} 
\Bigr ]^{1/2} \quad {\rm for} \quad q \in R_1
\label{wr4gonchainr1}
\eeq
and 
\beq
|W_r(\{(K_p \times (Ch)_4)_{rb}\},q)| = \Bigl | \Bigl [
1 -(2p+3)q^{-1}+(p^2+3p+3)q^{-2} \Bigr ] \Bigr |^{1/2} 
\quad {\rm for} \quad q \in R_2
\label{wr4gonchainr2}
\eeq

\subsection{$(K_p \times L_{n,bc})_{rb}$}

\begin{figure}
\epsfxsize=3.5in
\epsffile{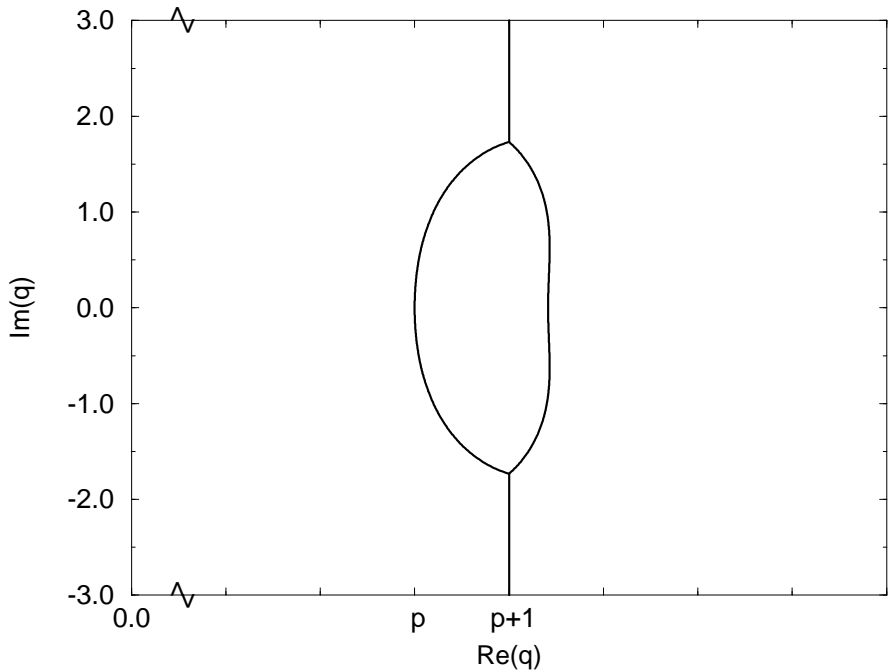}
\caption{Diagram showing regional boundaries comprising ${\cal B}$ for
$W(\{(K_p \times L_{xbc})_{rb}\},q)$ where $xbc$ denotes periodic or twisted
boundary conditions ($pbc$, $tbc$). \ Breaks in the horizontal axis indicate 
that $p$ is an arbitrary integer $\ge 2$.}
\label{kpladdercycfig}
\end{figure}

   A slightly more complicated illustration is provided by starting with an
$n$-vertex ladder graph, i.e., chain of squares, as in the $k=4$ case discussed
above, but with periodic or twisted boundary conditions (rather than open
boundary conditions) denoted $L_{n,pbc}$ and $L_{n,tbc}$, respectively.  
Note that any even number of twists is equivalent to no twist and any odd 
number of twists is equivalent to one twist.  Following our
general algorithm in (\ref{pkpcutb}), we adjoin $K_p$ to this ladder
graph and then remove $b$ of the bonds connecting a vertex in the $K_p$ graph
to other vertices in $K_p$. Using the basic results 
\beq
P(L_{n,pbc},q) = (q^2-3q+3)^{n/2} + 
(q-1)\Bigl \{ (3-q)^{n/2} + (1-q)^{n/2} \Bigr \} + q^2-3q+1
\label{ppbcladder}
\eeq
and
\beq
P(L_{n,tbc},q) = (q^2-3q+3)^{n/2} +
(q-1) \Bigl \{ (3-q)^{n/2} - (1-q)^{n/2} \Bigr \} - 1
\label{ptbclatter}
\eeq
we can apply our general analysis above.  First, we note that because the terms
raised to the $n$'th power, which can be leading terms as discussed in
conjunction with eq. (\ref{pgsum}), are the same for $P(L_{n,pbc},q)$ and 
$P(L_{n,tbc},q)$, it follows that 
\beq
W(\{(K_p \times L_{pbc})_{rb}\},q) = W(\{(K_p \times L_{tbc})_{rb}\},q)
\label{ptw}
\eeq
and hence the boundaries ${\cal B}$ are identical for these two functions. 
In both cases, from our general results, it follows that ${\cal B}$ 
contains complex-conjugate vertical line 
segments that extend to $\pm i \infty$ with
\beq
q_{_R}=p+1
\label{qrladder}
\eeq
as for the case of free boundary conditions, eq. (\ref{qr4gonchain}).  These 
line segments extend outward to $q_I = \pm \infty$ from the intersection 
points 
\beq
q_{int.}, \ q_{int.}^* = p+1 \pm i\sqrt{3}
\label{qintladder}
\eeq
At these intersection points, the boundary bifurcates into two, which extend
down and cross the real axis at 
\beq
q_{_{R, cross}} = p, \ p+\sqrt{2}
\label{qrcross}
\eeq
In Fig. \ref{kpladdercycfig} we show a plot of the region diagram for 
$W(\{(K_p \times L_{pbc})_{rb}\},q)=W(\{(K_p \times L_{tbc})_{rb}\},q)$; again,
by (\ref{bindb}), it is independent of $b$. 

\section{A General Condition Governing the (Non)analyticity of 
$W_{\lowercase{r}}(\{G\},\lowercase{q})$ at $1/\lowercase{q}=0$}

   As we stated at the end of section \ref{construction}, the key ingredient in
our algorithm to construct families of graphs $\{G\}$ with the property that 
${\cal B}$ extends to complex infinity in the $q$ plane and hence that $W_r(\{
G \},q)$ is nonanalytic at $1/q=0$ is to produce a chromatic polynomial with
the feature that there are two leading terms with a degeneracy condition
(\ref{mageq}).  From (\ref{pkpcutb}), these leading terms are actually the same
function, just evaluated at arguments that differ by unity, as is evident in
eq. (\ref{mageq}).  Here we generalize from this study to state the following
theorem: 

\vspace{2mm}

\begin{flushleft} 
Theorem. \ Consider a family of $n$-vertex graphs and its $n \to \infty$ limit
$\{H\}$.  With $P(H_n,q)$ in the form of eq. (\ref{pgsum}), 
the region boundary ${\cal B}$ of the resultant asymptotic limiting function 
$W(\{H\},q)$ extends to complex infinity in the $q$ plane, and hence
$W_r(\{H\},q)$ is nonanalytic at $1/q=0$ in the $1/q$ plane, if and only if the
locus of solutions of the degeneracy condition of leading terms
\beq
|a_\ell(q)| = |a_\ell'(q)|
\label{mageqgen}
\eeq
extends to complex infinity.  

\end{flushleft} 

\begin{flushleft}

Of course, this locus of points obeys the condition (\ref{binvariance}). 
Our algorithm yields the family of graphs $(K_p \times G_n)_{rb}$, depending on
$p$ and $b$, whose $n \to \infty$ limit, $\{(K_p \times G)_{rb}\}$
satisfies the above condition. 

\end{flushleft}

   A general feature of our results is that for the families of graphs we have
constructed and studied, for which ${\cal B}$ extends to complex infinity in
the $q$ plane, the image, under inversion, of ${\cal B}$ passes through the
origin of the $1/q$ plane with an infinite tangent, i.e., vertically.  This
reflects the property that the portion of ${\cal B}$ that extends to complex
infinity is comprised of a vertical line segment and its complex conjugate,
with a fixed value of $Re(q)$. 

  One salient feature of our study is clearly that none of our families of 
graphs with $W_r(\{G\},q)$ nonanalytic at $1/q=0$ is a regular 
lattice graph.  Our results are therefore consistent with the assumption
underlying the original series calculations, that a sufficient condition for 
$W_r(\{G\},q)$ to be analytic at $1/q=0$ is that $\{G\}$ be a regular lattice
graph $\Lambda$.  We state this formally as the following conjecture: 

\vspace{2mm}

\begin{flushleft}

Conjecture: 
Let $\{G\}$ denote the infinite-$n$ limit of a family of graphs $G_n$.  A
sufficient condition for the resultant reduced functions $W_r(\{G\},q)$
to be analytic at $1/q=0$ is that $\{G\}$ is a regular lattice graph 
$\{G\} = \Lambda$. 

\end{flushleft}

\vspace{2mm}

It is clear from Ref. \cite{w} that the property that $\{G\}$ be a regular
lattice is not a necessary condition for the associated $W_r(\{G\},q)$ to be
analytic at $1/q=0$; in that work we calculated $W(\{G\},q)$ functions for a
number of families $\{G\}$ which are not regular lattice graphs, but for which
the corresponding $W_r(\{G\},q)$ functions are analytic at $1/q=0$.

   For a regular lattice $\Lambda$ with coordination number $\zeta$, a 
natural reduced function is defined by 
\beq
W(\{G\},q)=q(1-q^{-1})^{\zeta/2}\overline W(\Lambda,y)
\label{wbar}
\eeq
for which the large-$q$ Taylor series expansion is \cite{nagle}-\cite{kewser} 
\beq
\overline W(\Lambda,y)=1+\sum_{n=1}^\infty w_{\Lambda,n} y^n \ , \quad 
y = \frac{1}{q-1}
\label{wbarseries}
\eeq
Clearly, $\overline W(\Lambda,y)$ is analytic at $y=0$ if and only if 
$W_r(\Lambda,q)$ is analytic at $1/q=0$. 
In Refs. \cite{w},\cite{ww},\cite{w3} we have made 
detailed comparisons of existing large-$q$ series expansions with 
high-precision Monte Carlo measurements of $W(\Lambda,q)$ as well as rigorous 
lower bounds that we have derived \cite{biggs} and have found excellent 
agreement for $q \ge 4$ on a number of different lattices, including both 
homopolygonal (e.g. square and honeycomb) lattices and heteropolygonal 
Archimedean lattices (composed of regular polygons of more than one type, 
such that all vertices are equivalent), viz., the $4 \cdot 8^2$ lattice, for
which we have calculated a large-$q$ series).  This excellent agreement
provides motivation to include lattices involving packings with different
regular polytopes under the term ``regular lattice'' in the above conjecture 
(here, polytope is the general mathematical object that subsumes the polygon 
in 2D and polyhedron in 3D \cite{coxeter}).

\section{Conclusions}

   In this paper we have addressed a fundamental problem in graph theory with
important implications for statistical mechanics, namely the question of the 
analyticity of $W_r(\{G\},q)$ at $1/q=0$.  In order to understand the 
phenomenon of nonanalyticity of this function at $1/q=0$ better, we have 
constructed a general algorithm for producing infinitely many families of 
graphs, each depending on two parameters $p$ and $b$, with $W_r$ functions that
are nonanalytic at $1/q=0$.  We have studied the properties of several of these
families. We have also stated a general necessary and sufficient condition on 
the chromatic polynomial of a family of graphs such that the resultant
$W_r(\{G\},q)$ is nonanalytic at $1/q=0$.  This condition explains the source 
of the nonanalyticity in the cases where it occurs. The results of our study 
are consistent with the conjecture that a sufficient condition (we know that
this is not a necessary condition) for $W_r(\{G\},q)$ to be analytic at 
$1/q=0$ is that $\{G\}=\Lambda$ is a regular lattice graph.  

This research was supported in part by the NSF grant PHY-93-09888.

\vfill
\eject

\begin{thebibliography}{99}
\bibitem{thm}{We recall the statement of the addition-contraction theorem 
\cite{rtrev}-\cite{biggsbook}: let $G$ be a graph, and let $v$ and $v'$ be two 
non-adjacent vertices in $G$.  Form (i) the graph $G_{add.}$ in which one 
adds a bond connecting $v$ and $v'$, and (ii) the graph $G_{contr.}$ in which 
one identifies $v$ and $v'$. Then the chromatic polynomial for coloring $G$ 
with $q$ colors, $P(G,q)$, satisfies $P(G,q) = P(G_{add.},q) + 
P(G_{contr.},q)$.}
\end{thebibliography}
\end{document}